\documentclass{emulateapj}

\usepackage{graphicx}
\usepackage{epstopdf}
\usepackage{color}
\pdfoutput=1

\def\gr{$\gamma$-ray}
\def\lsi{LSI~+61~303}
\def\fermi{FERMI}

\shorttitle{ LSI +61 303  X-ray superorbital modulation}
\shortauthors{Chernyakova et al.}

\begin{document}

\title{Superorbital modulation of X-ray emission from gamma-ray binary LSI +61 303}

\author{M.~Chernyakova}
\affil{Dublin City University, Glasnevin, Dublin 9, Ireland \\ DIAS, Fitzwiliam Place 31, Dublin 2, Ireland;}
%\and
\author{A.~Neronov}
\affil{ISDC Data Center for Astrophysics, Chemin d'Ecogia 16, 1290 Versoix,
Switzerland \\ Geneva Observatory, 51 ch. des Maillettes, CH-1290 Sauverny,
Switzerland}
%\and
\author{S.~Molkov}
\affil{Space Research Institute (IKI), 84/32 Profsoyuznaya Str, Moscow 117997, Russia}
\author{D.~Malyshev}
\affil{Bogolyubov Institute for Theoretical Physics,
14-b Metrolohichna street, Kiev, 03680, Ukraine}
%\and
\author{A.~Lutovinov}
\affil{Space Research Institute (IKI),84/32 Profsoyuznaya Str, Moscow 117997, Russia}
\and
\author{G.~Pooley}
\affil{Astrophysics, Cavendish Laboratory, Cambridge CB3 0HE, UK}

\begin{abstract}
We report the discovery of a systematic constant time lag between the X-ray and radio flares of the gamma-ray binary \lsi, persistent over long, multi-year, time scale. Using the data of monitoring of the system by RXTE we show that the orbital phase of X-ray flares from the source varies  from $\phi_X\simeq 0.35$ to $\phi_X\simeq 0.75$ on the superorbital 4.6~yr time scale.  Simultaneous radio observations show that periodic radio flares always lag the X-ray flare by $\Delta\phi_{X-R}\simeq 0.2$. We propose that the constant phase lag corresponds to the time of flight of the high-energy particle filled plasma blobs from inside the binary to the radio emission region at the distance $\sim 10$ times the binary separation distance. We put forward a hypothesis that the X-ray bursts correspond to the moments of formation of plasma blobs inside the binary system. 
\end{abstract}

\maketitle
%%%%%%%%%%%%%%%%%%%%%%%%%%%%%%%%%%%%%%%%%%%%

\section {Introduction.}

%%%%%%%%%%%%%%%%%%%%%%%%%%%%%%%%%%%%%%%%%%%%
\lsi\ is one of several high-mass X-ray binaries which emit  high-energy (GeV-TeV) \gr s (see e.g \cite{magic06,veritas08,fermi09}).  Although the nature of high-energy activity in high-mass X-ray binaries is not clearly understood,  a common hypothesis is that it is related to existence of  relativistic particle outflow.%, such as pulsar wind or a jet. This is definitely the case in PSR B1259-63 system, where \gr\ emission is produced in the collision of a pulsar wind with the wind from a massive star \citep{aharonian05,fermi_psrb}. In other \gr -loud binaries (GRLB), including \lsi, the nature of the outflow   is not clear yet. 

{\lsi\ is known to be variable on different time scales. The orbital period is $P_{orb}=26.496$~d \citep{gregory02}. Zero orbital phase $\phi=0$ corresponds to $T_{0,orb}=  2\,443\,366.775 + n P_{orb}$, periastron occurs at  $\phi=0.275$ \citep{aragona09}. The superorbital period is $P_{so}$=1667\,d \citep{gregory02}. The  zero superorbital phase $\Phi=0$ is $T_{0,so}=2\,443\,366.775+k P_{so}$.}
In \lsi\ the high-energy particle outflow is directly observed in the radio band, where angular resolution is sufficient to resolve the source and detect variations of its morphology on the orbital period time scale \citep{paredes98,massi04}. The observed morphological changes indicate that the outflow is, most probably, not a jet with a well defined position angle on the sky, but is rather a variable morphology outflow filling a region the size $\sim 10^2-10^3$ times larger than the binary separation distance \citep{dhawan06}. 

The radio signal could not be used to trace the  outflow   down to the production site inside the binary orbit, because of the free-free absorption in the dense stellar wind environment (e.g. \citep{zdziarski10}).  To understand the nature of the high-energy particle carrying  outflow one has to use  complementary high-energy data in X-ray and/or \gr\ bands.  

The X-ray and \gr\ emission from the system is known to be variable on the orbital time scale \citep{harrison00,chernyakova06,zhang10}. A study of the orbital modulation of the X-ray and \gr\ signal was missing up to recently due to the absence of systematic monitoring of the source on many orbit (year) time scale. Such a monitoring has recently became possible in the GeV \gr\ band with the start of operation of the Large Area Telescope (LAT) on board of Fermi satellite \citep{fermi09} which has now collected $\sim 3$~yr of data. In the X-ray band, a dedicated multi-year  monitoring campaign was done with RXTE \citep{smith09,torres10,li11}. As a result, both the average periodic modulation pattern and orbit-to-orbit variations of X-ray and \gr\ emission from the source were established.

In the X-ray band the source is known to exhibit one flare per orbit, on average preceeding the radio one.  The GeV band lightcurve during the first year of LAT observations also exhibited a one-flare pattern \citep{fermi09}. Both X-ray and GeV band lightcurves exhibit large orbit-to-orbit variations so that the systematic periodic variability is often washed out by an erratic behaviour of the source. The origin of the  X-ray and radio flares as well as the relation between the flaring activity of the source in different energy bands is not well understood.

 The orbital phase of the periodic flares drifts on a super-orbital time scale by half-an-orbit from $\phi_{r}\simeq 0.5$ to $\phi_r\simeq 1$ \citep{gregory02}. Such a drift is difficult to explain in scenarios based on various types of precession, where one expects a drift by a full orbit $0<\phi_r<1$ \citep{gregory02}. An alternative possibility for the explanation of the 4.6~yr time scale is the build up and decay of the equatorial disk around the massive Be star in the system \citep{zamanov99}. 

A new insight in the nature of the 4.6~yr periodicity/variability might be given by the study of the changes in the behaviour of the X-ray and \gr\ emission on this time scale. Here we report such a study based on the analysis of the monitoring of the source with RXTE, INTEGRAL and {\it Fermi}. The X-ray/\gr\ data are complemented by the contemporaneous radio monitoring data. 

%%%%%%%%%%%%%%%%%%%%%%%%%%%%%%%%%%%%%%%%%%%%
%\vskip0.1cm
\section {Data analysis.}
%%%%%%%%%%%%%%%%%%%%%%%%%%%%%%%%%%%%%%%%%%%%

In our analysis we consider radio, X-ray and \gr\ data on \lsi\ collected over the period from 2007 to 2011. 

LSI+61 303 was regularly observed by the PCA  \citep{bradt93}  instrument on board of RXTE 
(one $\sim 2-3$~ks observation every 3-5 days) since Sept. 2007 up to
Aug. 2011. During this period the set of operating detectors changed from observation to observation. Only detector number 2
(PCU2) operated all the time. For uniformity we consider only the data
from PCU2. To measure the source flux for an individual observation
we extracted an appropriate mean spectrum using the Standard-2 mode data. {For background subtraction we used the latest PCA background model for faint sources (see e.g. \cite{jagoda06} and PCA web-page). For the wide energy band (3-20 keV)
the backround estimation for pca2 is better than 0.1 mCrab. There are no known instrument related timescales close to orbital and superorbital periods.}
Most of the data analysis was performed using
the HEASOFT 6.11 software package.

\lsi\ was
several times in the field of view of INTEGRAL imager ISGRI   during the 
Galactic plane scans and pointed observations.  
In our analysis we  use all publicly available data for the period
from the January 2003 until October 2010. We process the data using the INTEGRAL Offline Science
Analysis (OSA  version 9.0) \citep{courvoisier}.

In the \gr\ band we use the data of Large Area Telescope (LAT) on board the \fermi\ satellite 
\citep{Atwood09} collected over the period from August 2008 to October 2011. We process the data using Fermi Science Tools  version 9r23p1 \footnote{http://fermi.gsfc.nasa.gov/ssc/data/analysis/}. The likelihood analysis is performed in a $10^\circ$ radius region around the source  taking into account all sources from the two-year LAT catalog \citep{fermi_catalog}.  Spectra of individual sources are modeled as powerlaws with the slope fixed to the best fit value from the two-year catalogue and free normalization.

The radio flux density was measured at 15.4\,GHz
with the Ryle Telescope and its successor, the 
AMI Large Array (Cambridge UK). The Ryle Telescope 
used a bandwidth of 750\,MHz, and the AMI uses 4\,GHz.
Pooley \& Fender (1997) describe the operation of the 
Ryle Telescope in observations of this type; a very 
similar technique is used for the AMI. Observations 
are typically 30 -- 60 min in duration.

%%%%%%%%%%%%%%%%%%%%%%%%%%%%%%%%%%%%%%
\begin{figure}
\includegraphics[width=\columnwidth,bb= 70 225 547 567,clip]{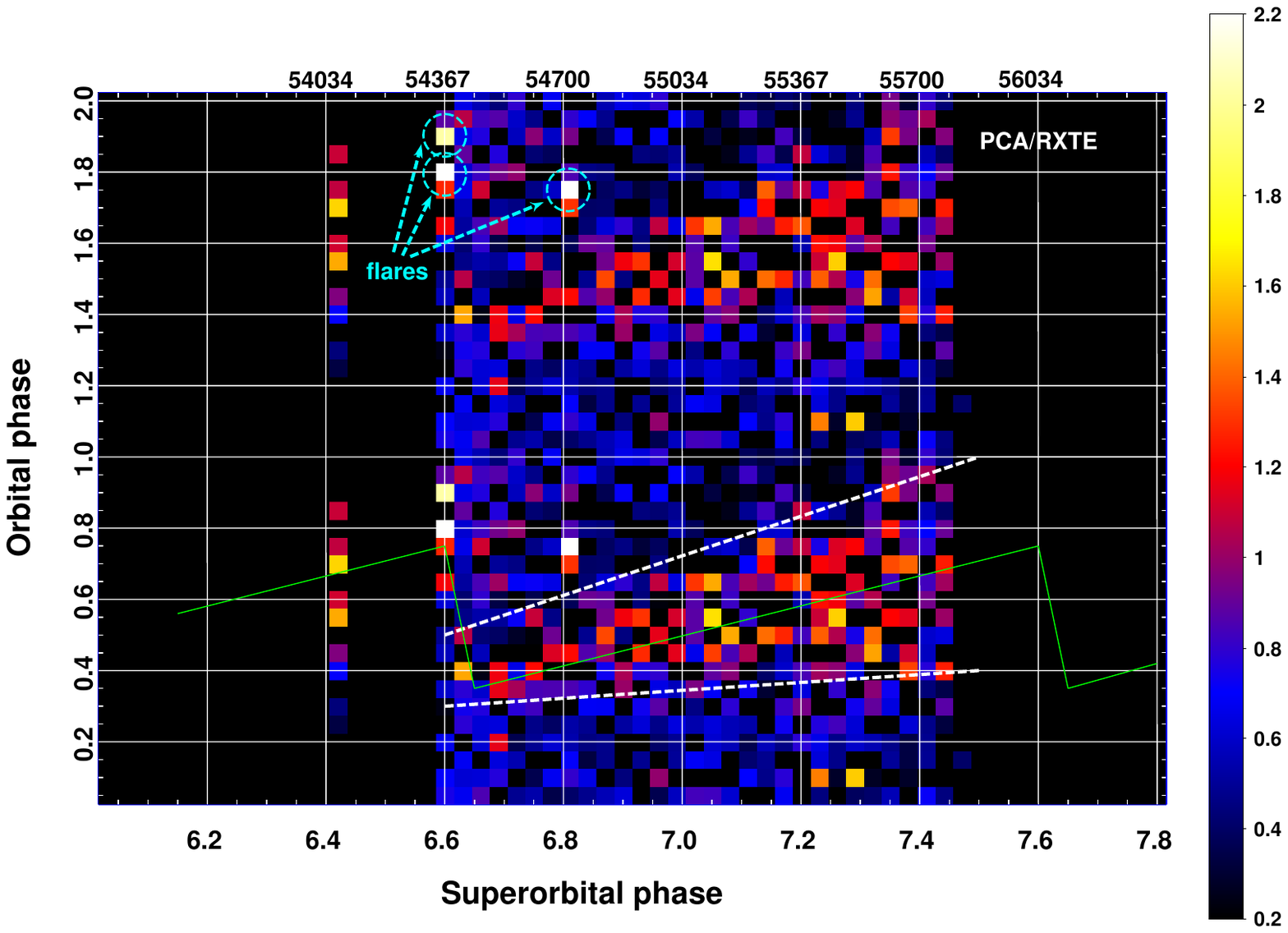}
\caption{3 -- 20 keV flux from \lsi\ as a function of the orbital vs. superobital phase. The color scale is expressed in mCrab units.}
\label{fig:pca}
\end{figure}
%%%%%%%%%%%%%%%%%%%%%%%%%%%%%%%%%%%%%%

%%%%%%%%%%%%%%%%%%%%%%%%%%%%%%%%%%%%%%%%%%%%
%\vskip0.1cm
\section{Results}
%%%%%%%%%%%%%%%%%%%%%%%%%%%%%%%%%%%%%%%%%%%%

Fig. \ref{fig:pca} shows the 3 -- 20 keV X-ray flux from the source as a function of the orbital and superorbital phase (as seen by RXTE/PCA). From Fig. \ref{fig:pca} one could see that the source exhibits on average one episode of increased X-ray activity per orbit. A regular increase of X-ray activity in the phase interval $\Delta \phi_X\simeq 0.2-0.4$ is accompanied by random variations of the source flux, with short flares appearing on the time scales $T_{flares,X}\ll \Delta\phi_XP_{orb}$. The average source flux in  3-20 keV energy range during the active/quiet part of the orbit indicated by the white dashed  lines on Fig. \ref{fig:pca} is about 1 mCrab/0.5 mCrab. Typical error of the flux measurment, $\sim$ 0.1 mCrab, is dominated by the uncertainity of the PCA background. The figure also has the color scale expressed in mCrab units. 

The average phase of the X-ray activity period $\phi_{X}$ varies on the superorbital time scale. From Fig. \ref{fig:pca} we find that $\phi_X$ exhibits a systematic drift from $\phi_X\simeq 0.35$ to $\phi_X\simeq 0.75$ within one superorbital cycle. Such a drift is similar to the systematic drift of the phase of the periodic radio flares from $\phi_R\simeq 0.5$ to $\phi_R\simeq 1$ \citep{gregory02}.

\begin{figure}
\includegraphics[width=\columnwidth,bb=24 420 585 675,clip]{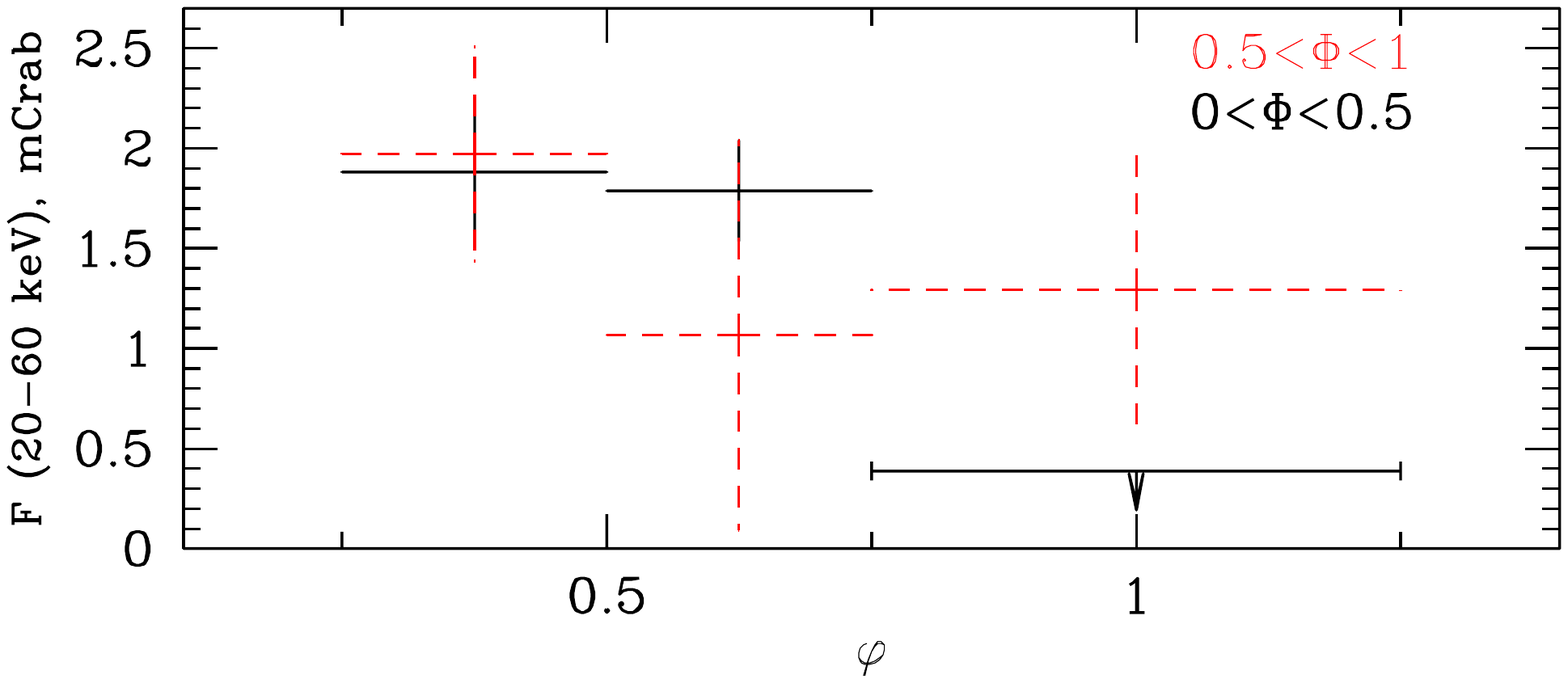}
\caption{Averaged orbital variability of the hard  X-ray flux (20 - 60 keV) from \lsi\ for the  $0<\Phi<0.5$ (black solid crosses) and  $0.5<\Phi<1$  (red dotted crosses) superobital phases. The corresponding exposures (from left to right) are 525, 664 and 452 ksec for the  $0<\Phi<0.5$ and 298, 132 and 207 ksec for the $0.5<\Phi<1$.}
\label{fig:int}
\end{figure}

Evolution of the orbital variability of the source in hard X-rays (20 - 60 keV) on the superorbital time scale, observed by INTEGRAL, is shown in Fig. \ref{fig:int}. One could see that, similarly to the 3-20~keV range,  the maximum flux happens during the orbital phase $0.25 <\phi < 0.5$ in the superorbital cycle phase $0.5<\Phi<1$. The maximum becomes wider and shifts toward $0.25 < \phi_X < 0.75$ in the superorbital phase $0<\Phi<0.5$.

%%%%%%%%%%%%%%%%%%%%%%%%%%%%%%%%%%%%%%
\begin{figure}
\includegraphics[width=\columnwidth,bb= 70 217 540 575,clip]{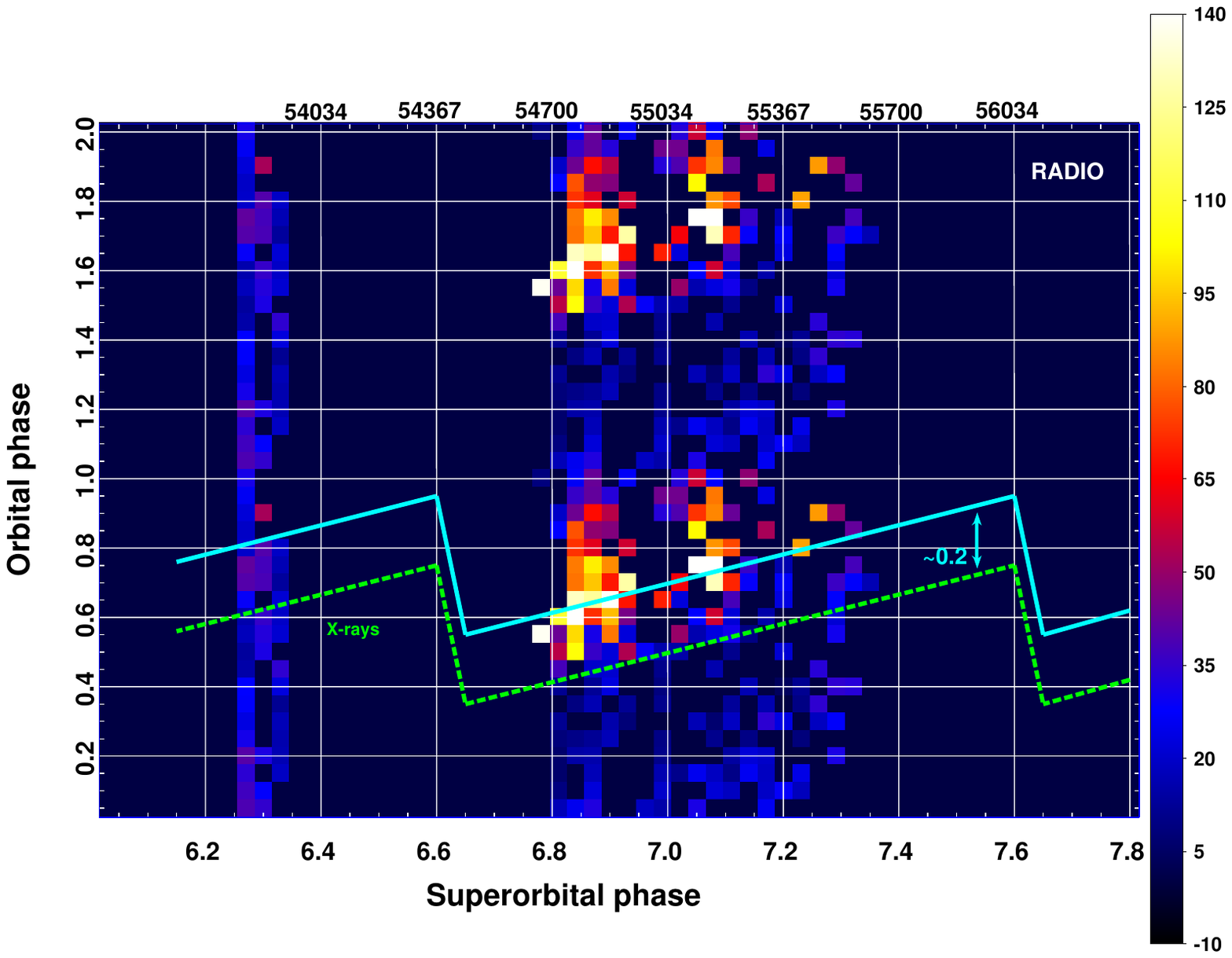}
\caption{Radio flux from \lsi\ as a function of the orbital and superorbital phases. The color scale is expressed in mJy units.}
\label{fig:radio}
\end{figure}

%%%%%%%%%%%%%%%%%%%%%%%%%%%%%%%%%%%%%%
Observations of the systematic drift of the phase of the radio flares from the source reported by \cite{gregory02} were performed several superorbital cycles before the X-ray monitoring campaign by RXTE. To verify the long-term stability of the range of the shifts of $\phi_R$ over many superorbital cycles we use the data of monitoring of the source in the radio band which are contemporaneous with the RXTE monitoring campaign. Fig. \ref{fig:radio} shows the radio flux of the source as a function of the orbital and superorbital phases. {Colorbar shows flux measured in mJy.} 
Comparing Fig. \ref{fig:radio} with the equivalent figure from \citet{gregory02}, we find that the overall drift pattern of the phase of the radio flare remained stable over several superorbital cycles. The same drift from the phase $\phi_R\simeq 0.55$ to $\phi_R\simeq 0.95$ is observed also in the radio data contemporaneous with the RXTE monitoring campaign.

%%%%%%%%%%%%%%%%%%%%%%%%%%%%%%%%%%%%%%
%%%%%%%%%%%%%%%%%%%%%%%%%%%%%%%%%%%%%%

Comparison of X-ray and radio superorbital variability patterns is shown in Figure \ref{fig:radio}. The average phase of the X-ray activity period always preceeds the phase of the radio flare by $\Delta\phi_{X-R}\simeq 0.2$, which corresponds to the time delay $\Delta T_{X-R}=\Delta\phi_{X-R}P\simeq 5.3$~d. 

\begin{figure}
\includegraphics[width=\columnwidth,bb= 60 220 555 575,clip]{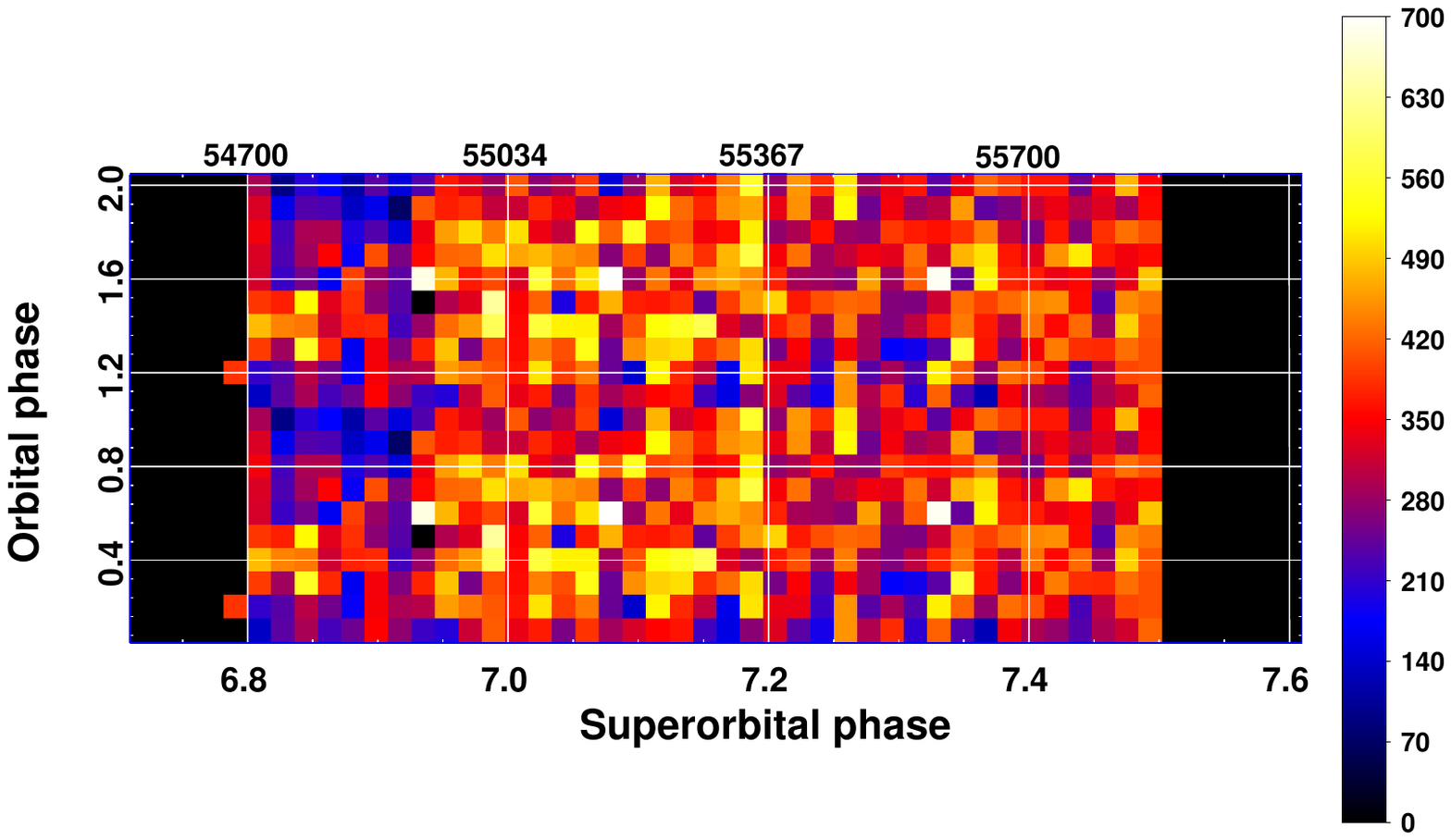}
\caption{Very high energy (E$>$100 MeV)  flux from \lsi\ as a function of the orbital and superorbital phases. The color scale is expressed in mCrab units.}
\label{fig:fermi}
\end{figure}

Contrary to the X-ray and radio bands, the superorbital modulation pattern is not clearly visible in the \gr\ band.  Fig. \ref{fig:fermi} shows the source flux in the 0.1-10~GeV energy band plotted as a function of the orbital and superorbital phase, similarly to Figs. \ref{fig:pca} and \ref{fig:radio}. {Colorbar shows flux measured in mCrabs, typical error is about 10\% of the flux.} Long-term source behaviour of the source in the GeV band is puzzling.  Orbital modulation was clearly observable at the beginning of Fermi observations at the superorbital phase $6.8<\Phi<6.9$. The phase of the maximum orbital modulation of the GeV flux in this superorbital phase range was close to the phase of the X-ray activity. However, in the time period following the superorbital phase $\Phi\simeq 6.9$ a clear orbital modulation pattern disappeared (see also \cite{hadasch11}). Further study of the source on the time scale of several superorbital cycles is needed to clarify the repeatability of the observed appearance / disappearance of the orbital modulation pattern and its relation to the overall 4.6~yr activity cycle of the source.

%%%%%%%%%%%%%%%%%%%%%%%%%%%%%%%%%%%%%%
%%%%%%%%%%%%%%%%%%%%%%%%%%%%%%%%%%%%%%

%\vskip0.1cm
\section{Discussion.} 

A constant time delay between the drifting orbital phases of X-ray and radio flares could be naturally explained if  one takes into account that radio and X-ray emission originate from different regions.  The radio emission is produced at large distance from the binary, $D_R\gtrsim 5\times 10^{13}$~cm \citep{zdziarski10}  while the X-ray flux is most probably produced at shorter distances of the order of the binary separation $3\times 10^{12}\mbox{ cm}<D_X <10^{13}\mbox{ cm}$. Assuming that injection of high-energy electrons responsible for the X-ray and radio flares happens in the same event in the binary, one could attribute the time delay between the X-ray and radio flares to the time-of-flight of the high-energy particle filled plasma to the radio emission region. {The cooling times of high energy particles are long enough to allow them to travel to the radio emission region.} This gives an estimate of the plasma outflow velocity $v_R\simeq D_R/\Delta T_{X-R}\simeq 10^8$~cm/s, which is in good agreement with the asymptotic velocity of the stellar wind from the massive star  in both the polar and equatorial regions \citep{zdziarski10}. 

The constant time delay between the X-ray and radio flare phases suggests the following scenario of production of the periodic radio flares. Once per orbit, an event of interaction of the compact object with the stellar wind leads to injection of high-energy particles into the stellar wind. The high-energy particles mixed into the stellar wind escape from the binary system with the stellar wind velocity. A radio flare occurs at the moment when the portion of the stellar wind filled with high-energy electrons reaches the distances $D\sim D_R$ at which the system becomes transparent to the radio waves. 

High-energy electrons, responsible for the radio-to-X-ray emission, are held in the plasma outflow by the magnetic field. Electrons producing synchrotron emission in the radio band have energies  $E_e\simeq 10 \left[B/1\mbox{ G}\right]^{-1/2}\left[\nu_R/10\mbox{ GHz} \right]^{1/2}$~MeV, where $\nu_R$ is the frequency of the radio synchrotron emission. The main cooling mechanisms for such electrons are synchrotron and/or  inverse Compton emission with the characteristic cooling time scales $t_S\simeq 0.7 \left[B/1\mbox{ G}\right]^{-3/2}\left[\nu_R/10\mbox{ GHz}\right]^{-1/2}$~yr and $t_{IC}\simeq 6.5\left[D/10^{13}\mbox{ cm}\right]^{-2}\mbox{ d}$. Unless magnetic field in the system is much higher than $B\sim 10$~G, the synchrotron and inverse Compton cooling time scales are not shorter than the time-of-flight from inside the binary orbit to the radio emission region. Thus energy losses do not prevent electrons injected in the binary system from traveling to the radio emission region on a time scale of several days.  

The X-ray emission is most probably produced via the synchrotron mechanism by electrons of the energies $E_e\simeq 100\left[B/10\mbox{ G}\right]^{-1/2}\left[E_X/5.4\mbox{ keV}\right]^{1/2}\mbox{ GeV}$. This conjecture is supported by the observations of fast variability of X-ray emission on the time scales $t\sim 10$~s \citep{smith09} which is comparable to the synchrotron cooling time of 100~GeV electrons in the $B\sim 10$~G magnetic field. The absence of any obvious break / cut-off features in the keV-GeV source spectrum with a maximum in the $\sim 10-100$~MeV range is  in favour of interpretation of the entire keV-GeV bump as a single spectral component. 
In such a model the GeV band emission is due to the synchrotron emission by the 10-100~TeV electrons.

The phase of the \gr\ flare is close to the phase of the X-ray flare (Fig. \ref{fig:pca} and \ref{fig:fermi}), at least during a part of the superorbital cycle $6.8<\Phi<6.9$. It is natural to identify the phase of the X-ray/\gr\ flare with the moment of formation of high-energy particle outflow inside the binary. Long term evolution of this phase could not be followed in the \gr\ band, because of the large width of the activity period after the superorbital phase $\Phi>6.9$. In the X-ray band the width remains finite  all over the superorbital cycle. Difference in the superorbital modulation pattern in X-rays and \gr s is, most probably, related to different regimes of acceleration/propagation/cooling of 10 MeV and 10 TeV electrons.

The phase of the X-ray  activity is close to the phase of the periastron of the binary orbit $\phi_{per}\simeq 0.3$ at the superorbital phase $\Phi\simeq 0.5$ (Fig. \ref{fig:pca}) when the duration of the activity period is shortest. 
The phase of activity gradually shifts to the post-periastron interval $\phi_X>0.3$ over the superorbital period and almost reaches the phase of the apastron $\phi_{ap}\simeq 0.8$ toward the end of the superorbital cycle so that the X-ray emission is always delayed with respect to the apastron. Shift of $\phi_X$ is accompanied by the increase of the width of activity period. 

A possible explanation for such behaviour could be found in a scenario in which the 4.6~yr superorbital cycle is interpreted as the cycle of gradual buildup and decay of the equatorial disk of the Be star. At $\Phi\simeq 0.5$ the equatorial disk is weak. The phase of the closest encounter between the disk and the compact object is the phase of the periastron. Interaction of the compact object with the disk perturbs the disk and strips away a part of the disk which escapes from the system to the radio emission region. Gradual buildup of the equatorial disk due to ejection of matter from the Be star leads to the increase of the disk density and/or disk size. {Such a scenario implies that extended radio emission shouldn't have a clear jet-like morphology, but rather have an irregular morphology varying over the orbital and superorbital cycle.Such a variable morphologu is indeed observed in the radio band \cite{dhawan06}.}

The shift of the phase $\phi_X$ of ejection of a portion of the disk could be explained by the increase of the time of  accumulation of energy sufficient for ejection. The kinetic energy needed to strip away a part of the disk is comparable to the gravitational binding energy of the disk, $U\sim G_NM_*\rho_dR_dH_d\sim 10^{40}\left[\rho_d/10^{13} \mbox{g} \cdot \mbox{cm}^{-3}\right]\left[R_d/10^{12}\mbox{ cm}\right]\left[H_d/10^{12}\mbox{ cm}\right]\times$ $[M_*/10M_\odot]$~erg, where $\rho_d, R_d$ and $H_d$ are the density, radius and thickness of the equatorial disk. Such energy should be transmitted to the disk by the compact object at each disk-compact object interaction event. Suppose that the compact object injects energy in the disk at a constant rate $P$~erg/s at each interaction event. The energy sufficient for ejection of a part of the disk is then accumulated on a time scale $T_{ej}\sim U/P\sim 1\left[P/10^{35}\mbox{ erg/s}\right]^{-1}$~d. Accumulation of mass in the disk leads to the increase of $T_{ej}$ and, as a consequence, to the shift of $\phi_X$.  

If the disk  size reaches the size of the binary orbit,  the compact object always moves inside the disk and continuously perturbs it. Studies of the high-mass X-ray binaries with Be stars show that in this case the compact object induced instabilities in the disk might lead to destruction and 
complete loss of the disk. Our hypothesis is that the loss of the disk corresponds to the period in the superorbital phase range $\Phi\simeq 0.4-0.5$ when  the strength of X-ray and radio flares decreases and no systematic periodic variability of the source is observed (see Fig. \ref{fig:pca}, \ref{fig:radio} and \citet{gregory02}). Ejection of matter from the equatorial regions of Be star leads to formation of a new disk at around $\Phi\simeq 0.5$ and a new cycle of disk growth/decay starts. 

In such a scenario, regularity of the superorbital modulation in the system could be readily explained. The constant growth rate of the equatorial disk of Be star is determined by the stable rotation of the Be star. The period of superorbital modulation is determined by the fixed time scale on which the disk growth to the size comparable to the size of the binary orbit. This scenario for the origin of the orbital and superorbital modulations of the source flux could be tested using the H$\alpha$ data which provide a diagnostic of the state of the equatorial disk of Be star \citep{zamanov99,mcswain10}. Growth and decay cycles of the disk  lead to the variations of the overall strength and shape of the H$\alpha$ line. Variability of the line intensity and profile on the time scale of the superorbital modulation was demonstrated by \citet{zamanov99}. Monitoring of the H$\alpha$ line on the time scale of several superorbital cycles would show if the the observed variability corresponds to the periodic buildup and decay of the disk. Periodic ejection of a part of the disk as a result of the compact object-disk interaction might be responsible for occurence of transient red or blue ``shoulders'' of H$\alpha$ line as observed by \citet{mcswain10}. Systematic re-observation of the repetition of occurence of the shoulders in many orbital cycles and correlation of the phase of occurence of the shoulders with the phases of X-ray flares would provide a direct test for our model.

\textit{Note added in proof:} During the process of publication of this
article, a similar study by \cite{li12} appeared in press.

\textbf{Acknowledgements}
The authors thank participants of the ISSI team ``Study of Gamma-ray Loud Binary Systems'' for useful discussions, and the International
Space Science Institute (ISSI, Bern) for support. The authors also wish to acknowledge the SFI/HEA Irish Centre for High-End Computing (ICHEC) for the provision of computational facilities and support. The work of D.M. is supported in part by the Cosmomicrophysics programme of the National Academy of Sciences of Ukraine and by the State Programme of Implementation of Grid
Technology in Ukraine. S.M. and A.L. acknowledged the support from the program ``Origin, Structure and Evolution of the Objects in the Universe'' by the Presidium of the Russian Academy of Sciences, grant no.NSh-5069.2010.2 from the President of Russia, Russian Foundation for Basic Research (grants 11-02-01328 and 11-02-12285-ofi-m-2011), State contract 14.740.11.0611

\end{document}